\documentclass[11pt,a4paper]{article}

\usepackage{geometry}
\usepackage{setspace}

\usepackage{bbm}
\usepackage{graphics} 
\usepackage{epsfig} 
\usepackage{mathptmx} 
\usepackage{times} 
\usepackage{amsmath} 
\usepackage{amssymb}  
\usepackage{amsfonts}
\usepackage[centerlast]{caption}
\usepackage{graphicx}          

\usepackage{tabularx}
\usepackage{booktabs}

\usepackage{amssymb}
\newtheorem{theorem}{Theorem}
\newtheorem{proposition}{Proposition}
\newtheorem{definition}{Definition}

\newtheorem{corollary}{Corollary}

\newtheorem{remark}{Remark}
\newtheorem{example}{Example}

\newtheorem{problem}{Problem}

\usepackage{amsmath}

\begin{document}

\title{Leader Selection and Weight Adjustment for Controllability of Multi-Agent Systems}

\author{Bin Zhao
\thanks{Bin Zhao and Long Wang are with Center for Systems and Control, College of Engineering, Peking University, Beijing,
100871, China e-mail: bigbin@pku.edu.cn, longwang@pku.edu.cn.},
        Yongqiang Guan
\thanks{Yongqiang Guan is with College of Automation Science and Electrical Engineering, Beihang University, Beijing, 100191, China
                  e-mail: guan-jq@163.com.},
        and Long Wang$^*$}

\maketitle

\begin{abstract}
For an uncontrollable system, adding leaders and adjusting edge weights are two methods to improve controllability. In this paper, controllability of multi-agent systems under directed topologies is studied, especially on leader selection problem and weight adjustment problem. For a given system, necessary and sufficient algebraic conditions for controllability with fewest leaders are proposed. From another perspective, when leaders are fixed, controllability could be improved by adjusting edge weights, and therefore the system is supposed to be structurally controllable, which holds if and only if the communication topology contains a spanning tree. It is also proved that the number of fewest edges needed to be assigned on new weights equals the rank deficiency of controllability matrix. An algorithm on how to perform weight adjustment is presented. Simulation examples are provided to illustrate the theoretical results.\\
\textbf{keywords:} Multi-agent systems; Controllability; Leader selection; Weight adjustment
\end{abstract}

\section{Introduction}

In the past few decades, due to the rapid development of computer science and communication technology, distributed cooperative control of multi-agent systems has become a hot topic in multidisciplinary research area. Many results have been obtained and applied in science and engineering areas, such as flocking in biology, formation of unmanned air vehicles, attitude alignment of satellite clusters and data fusion of sensors, etc. Researches on multi-agent systems include several fundamental problems, such as consensus \cite{Xiao07}, formation \cite{Xiao09}, flocking and swarming \cite{Saber06,Shi04}, stabilizability \cite{Guan13,Guan14} and controllability \cite{Tanner04}, etc.

Controllability is a significant issue on multi-agent systems and attracts increasing attentions. A multi-agent systems is said to be controllable if appropriate external controls are put on the leaders such that all agents will achieve any designed configuration from any given initial states within a finite time. The controllability problem of multi-agent systems was put forward for the first time by Tanner \cite{Tanner04}, where an algebraic necessary and sufficient condition was presented under undirected communication topologies. Based on this, Ji et al. proposed a leader-follower connected structure and proved it to be a necessary condition to control a multi-agent system with multiple leaders \cite{Ji08,Ji09}. The models of agents used in the above are all with single-integrator dynamics. In \cite{Wang09}, Wang et al. studied systems whose agents are with high-order dynamics and generic linear dynamics, and proved that controllability is congruously determined by the communication topology, regardless of agents' dynamics. Further researches presented necessary and sufficient conditions for controllability on some special graphs, such as cycles and paths \cite{Parlangeli12}, stars and trees \cite{Ji12}, grid graphs \cite{Notarstefano13} and regular graphs \cite{Kibangou}, to name a few. Conclusions on directed topologies are only confined to strongly regular graphs and distance regular graphs \cite{Kibangou}, graph partitions \cite{Lou12} and some specific graphs \cite{Guan}. With respect to switching topologies, Liu et al. achieved several results on controllability \cite{Liu08,Liu12}.

A parallel research line in this field is structural controllability, which was proposed by Lin in \cite{Lin74} for linear time-invariant systems, and was brought into multi-agent systems in \cite{Zamani009}. On the one hand, structural controllability was investigated under various models \cite{Lin74,Zamani009,Mayeda79}, whereas all conclusions in \cite{Zamani009} ignored the $0$ row sum restriction of the Laplacian matrix. In other words, the interactions of the agents were not based on distributed consensus protocol. Although the protocol in \cite{Lou12} and \cite{Liu13} is a distributed one, the structurally controllable problem proposed there either allows adding edges between some agents, or is focused on undirected graphs. On the other hand, the existing results only qualitatively judged the structural controllability. However, if a system is not strongly structurally controllable \cite{Mayeda79}, how to arrange a set of feasible weights that could ensure controllability is critically important. Moreover, for an uncontrollable system, in the premise of not adding any leader, controllability could be also improved by adjusting edge weights. To our best knowledge, how to adjust the weight parameters in a multi-agent system, especially how to choose the fewest edges and endow proper new weights to improve controllability has not been studied.

Inspired by previous results, this paper studies leader selection problem and weight adjustment problem for multi-agent systems under directed communication topologies. The contributions in our research are threefold: (i) All conclusions in this paper are based on directed weighted graphs. (ii) Necessary and sufficient algebraic conditions of single leader controllability and fewest leaders to control a system are provided, via the Jordan form of Laplacian matrix and corresponding similarity transformation matrix. Jordan form of Laplacian matrix becomes a new channel to investigate controllability of multi-agent systems. (iii) Necessary and sufficient graphic conditions for structural controllability are firstly given under the distributed consensus protocol; a new problem named ``weight adjustment'' is put forward, which solves that how to get the exact fewest edges to be assigned new weights to achieve controllability quantitatively, along with the algorithm of proceeding weight adjustment.

This paper is organized as follows: In Section 2, basic concepts and preliminaries are given. In Section 3, leaders selection problem is investigated. In Section 4, weight adjustment problem is proposed and solved. An application on checking controllability of in-degree regular graphs is shown in Section 5. Two typical examples are shown in Section 6 to illustrate the theoretical results. Finally, the conclusions are summarized in Section 7.

$\mathbf{Notations:}$ Throughout this paper, the following notations are used. $\mathbf{1}_n$ is a vector with dimension $n$ whose entries are all $1$, and sometimes footprint $n$ is omitted for convenience. If $A$ is a square matrix, $\det(A)$ denotes the determinant of $A$. $diag(a_1,a_2,\cdots,a_n)$ and $\max\{b_1,b_2,\cdots,b_m\}$ represent the diagonal matrix with principal diagonals $a_1,a_2,\cdots,a_n$ and the maximum value in $b_1,b_2,\cdots,b_m$, respectively. The set of all real numbers is denoted by $\mathbb{R}^n$.
A vector $\alpha$ is called ``all-$0$'' if all the entries in $\alpha$ are $0$. A matrix is called ``all-$0$'' if all of its columns are all-$0$. $\Lambda(A)$ denotes the eigenvalue set of $A$. $|S|$ represents the cardinality of a set $S$.

\section{Preliminaries and problem formulation}

\subsection{Graph theory}

A directed graph $\mathbb{G}=(\mathbb{V},\mathbb{E})$ consists of two parts, $\mathbb{V}=\{v_1,v_2,\cdots,v_n\}$ is the set of nodes in the graph, and $\mathbb{E}\subseteq \mathbb{V}\times\mathbb{V}$ represents the edge set. An edge in $\mathbb{E}$ is denoted by $(v_i,v_j)$ if the edge points at $v_j$ from $v_i$. $v_i$ is called the parent node while $v_j$ is called the child node and we say $v_i$ is a neighbor of $v_j$. The neighbor set of $v_j$ is denoted by $N_j=\{v_i\in\mathbb{V}|(v_i,v_j)\in\mathbb{E}\}$. The in-degree of node $v$ is the total number of its neighbors, denoted as $\deg_{in}(v)$. Assume that there is no self-loop at any node, i.e. $(v_i,v_i)\notin\mathbb{E}$, hence not any node is a neighbor of itself. A directed path $\mathbb{P}_n$ is a graph with $n$ nodes and the edges are only $(v_1,v_2),(v_2,v_3),\cdots,(v_{n-1},v_n)$. A tree graph $\mathbb{T}_v$ with root $v$ is a graph that for each node other than $v$, there exists one and only one path from $v$ to this node. In a tree graph, a node is called a leaf if it has no child, and two nodes are said to be in different branches when there is no path from any one of them to the another. A graph $\mathbb{G}$ is said to contain a spanning tree if there exists a tree whose nodes are all those in $\mathbb{V}$ and edges in the tree are also in $\mathbb{E}$. A spanning forest of $\mathbb{G}$ is a set of trees covering $\mathbb{V}$ with no common nodes, and edges are all in $\mathbb{E}$. The minimal spanning forest is a spanning forest with fewest trees.
Length of the shortest path from $v_i$ to $v_j$ is called the distance from $v_i$ to $v_j$, denoted by $d(v_i\rightarrow v_j)$. Especially, if $v_i=v_j$, $d(v_i\rightarrow v_j)=0$. $d(v_i\rightarrow v_j)=\infty$ when there is no path from $v_i$ to $v_j$. The distance partition is defined as follow.
\begin{definition}
The distance partition of graph $\mathbb{G}$ relative to node $v$ consists of a series of sets $D_0,D_1,D_2,\cdots,D_l$ and $D_\infty$, where $D_0=\{v\}$, $D_i=\{w\in \mathbb{V}|d(v\rightarrow w)=i\}$ and $D_\infty=\{w\in \mathbb{V}|d(v\rightarrow w)=\infty\}$. $\bigcup\limits_i D_i=\mathbb{V},i=0,1,2,\cdots,l,\infty$.
\end{definition}
In this paper, $\mathbb{G}$ is fixed. The adjacency matrix of $\mathbb{G}$ is $A(\mathbb{G})=[a_{ij}]\in \mathbb{R}^{n\times n}$, where $a_{ij}$ is the weight of edge $e_{ij}$, and $a_{ij}=0$ if $(v_j,v_i)\notin\mathbb{E}$. The Laplacian matrix of $\mathbb{G}$ is $L=D-A$, $D=diag(d_1,d_2,\cdots,d_n)$ where $d_k=\deg_{in}(k)$ is the in-degree of node $k,k=1,2,\cdots,n$. A matrix $M$ is said to be $\emph{cyclic}$ if its eigenpolynomial equals the minimal polynomial.

Since the mapping between the communication topology of a system and the corresponding graph is a bijection, ``node'' and ``agent'' are not distinguished in this paper for convenience.
\subsection{Problem formulation}
Consider a multi-agent system with $n$ single-integrator dynamic agents:
\begin{equation}\label{model}
\dot{x}_i=u_i,~i=1,2,\cdots,n.
\end{equation}
Here $x_i$ and $u_i$ represent the state and the control input on agent $i$. For simplicity, only one dimensional states are considered in the following, i.e. $x_i\in \mathbb{R}$. However, the results obtained from this paper can be extended to arbitrary dimensional systems via Kronecker products. Agents that can be driven by external inputs are called $\mathbf{leaders}$. The set of leaders are denoted by $\mathbb{V}_l=\{i_1,i_2,\cdots,i_m\}$. The rest agents are called $\mathbf{followers}$, denoted by $\mathbb{V}_f=\mathbb{V}/\mathbb{V}_l$. The control inputs on the agents obey a distributed consensus-based protocol:
\begin{eqnarray}\label{protocol}
u_i = \left\{
\begin{aligned}
   &{\sum\limits_{j \in N_i } ( x_j  - x_i ) + u_{o,i} ,} &&{i \in \mathbb{V}_l } , \\
   &{\sum\limits_{j \in N_i } ( x_j  - x_i ),} &&{i \in \mathbb{V}_f },\\
\end{aligned}\right.
\end{eqnarray}
where $u_{o,i}$ is the external control on agent $i$.

The compact form of system (\ref{model}) with protocol (\ref{protocol}) is summarized as follows.
\begin{equation}\label{compact}
  \dot{x}=-Lx+Bu,
\end{equation}
where $x=(x_1,x_2,\cdots,x_n)^T\in\mathbb{R}^n$ and $u=(u_1,u_2,\cdots,u_m)^T\in\mathbb{R}^m$ represent the states and control inputs, respectively. $L$ is the Laplacian matrix and $B=(e_{i_1},e_{i_2},\cdots,e_{i_m})\in\mathbb{R}^{n\times m}$. $e_i\in\mathbb{R}^n$ is a vector with the $i$-th entry $1$ and the rest $0$. \begin{definition}\label{controllale}
Multi-agent system (\ref{compact}) (or the corresponding communication graph) is said to be controllable if for any initial state $x(t_{0})$ and target state $x^*$, $x(t_{0})$ can be actuated to $x(t_1)=x^*$ in finite time $t_1>t_0$ with external controls $u$ on leaders.
\end{definition}
Especially, if all $0$ entries in the adjacency matrix of the communication graph remain to be $0$, and all other entries can be weighted positive numbers freely, the concept of structural controllability is proposed.
\begin{definition}\label{structural}
Multi-agent system (\ref{compact}) (or the corresponding communication graph) is said to be structurally controllable if there exists a group of weights to make the system controllable.
\end{definition}

If multi-agent system (\ref{compact}) is not controllable, there are two methods to improve controllability, i.e. adding leaders or adjusting edge weights. The former one derives leader selection problem, which will be introduced in the next section, and the latter method derives weight adjustment problem, which will be discussed in Section 4.

\section{Leader selection problem}

\begin{problem}\label{lsp}
$\mathbf{Leader~selection~problem}$: For multi-agent system (\ref{compact}), find a set of nodes $\mathbb{V}_l\subseteq\mathbb{V}$ with minimum $|\mathbb{V}_l|$, such that when all nodes in $\mathbb{V}_l$ are chosen as leaders, the system is controllable.
\end{problem}
The investigation of leader selection problem begins with a basic concept.
\begin{definition}\label{lcs}
$r$ Leaders Controllable System: Multi-agent system (\ref{compact}) is said to be $r$ leaders controllable if the minimum $|\mathbb{V}_l|=r$. Especially, if $r=1$, system (\ref{compact}) is called single leader controllable (SLC).
\end{definition}
\subsection{Single leader controllability}

The leader selection problem is started by single leader controllability. As a matter of fact, controllability of system (\ref{compact}) is invariant under any labeling of the nodes in communication graph $\mathbb{G}$. Suppose $B=e\triangleq e_i\in\mathbb{R}^n$. Since the controllability of system (\ref{compact}) is same to that of system $\dot{x}=Lx+Bu$, the latter system is studied for simplicity. Consider the controllability matrix $C = (e, Le,{L^2}e, \cdots , L^{n - 1}e)$, system (\ref{compact}) is controllable if and only if $rank(C)=n$.

Denote the Jordan form of $L$ as $J=diag(J_0,J_1,J_2,\cdots,J_s)$, and the corresponding similarity transformation matrix is $P = ({\xi _1},{\xi _2},{\xi _3}, \cdots ,{\xi _n})$, $P^{-1}LP=J$. Here
\begin{equation*}
{J_j} = {\left( {\begin{array}{*{20}{c}}
{{\lambda _j}}&1&{}&{}\\
{}&{{\lambda _j}}& \ddots &{}\\
{}&{}& \ddots &1\\
{}&{}&{}&{{\lambda _j}}
\end{array}} \right)_{{n_j} \times {n_,}}}, j=0,1,\cdots,s, n_0+n_1+n_2+\cdots+n_s=n.
\end{equation*}
Denote $m_j=n_0+n_1+n_2+\cdots+n_j$, $j=1,2,\cdots,s$, obviously $m_s=n$. $\xi_{1}, \xi_{m_0+1}, \xi_{m_1+1}, \cdots, \xi_{m_{s-1}+1}$ are the linearly independent eigenvectors of $L$. Correspondingly, $\xi_{m_j+2},\xi_{m_{j}+3},\cdots,\xi_{m_{j+1}}$ are the linearly independent generalized eigenvectors of $\lambda_j$.

\begin{theorem}\label{slc}
Multi-agent system (\ref{compact}) is SLC if and only if the following two conditions are satisfied simultaneously:\\
1. The Laplacian matrix $L$ is cyclic;\\
2. There exists a column $\eta^i$ in $P^{-1}= (\eta^1,\eta^2,\cdots,\eta^n)$ such that $\eta_{m_j}^i\neq 0$ for all $j=0,1,2,\cdots,s$.\\
In this circumstance, agent $i$ is able to control the system, where $i$ is also the column index of $\eta^i$ in $P^{-1}$.
\end{theorem}
Proof: According to the PBH Test, system (\ref{compact}) is controllable if and only if $rank(\lambda I+L,B)=n$ for any $\lambda\in\Lambda(-L)$. Since $rank(\lambda I+L,B)=rank(\lambda I+J,P^{-1}e)$, consider rank of $(\lambda I+J,P^{-1}e)$. If there are two different Jordan blocks in $J$ sharing the same eigenvalue, the two rows in $(\lambda I+J,P^{-1}e)$ corresponding to the last rows of the two Jordan blocks will always be linearly dependent, which means $rank(\lambda I+J,P^{-1}e)<n$, therefore condition 1 is necessary. When the Laplacian matrix $L$ is cyclic, all Jordan blocks in $J$ have different eigenvalues, thus $rank(\lambda I+J,P^{-1}e)=n$ for all $\lambda\in\Lambda(-L)$ if and only if the $m_0$-th, $m_1$-th, $m_2$-th, $\cdots$, $m_s$-th entries of $P^{-1}e$ all not be $0$. If $e=e_i$ could satisfy this condition, the system is controllable, and meanwhile, only the dynamic of the $i$-th agent is affected by the external input $u$, i.e. agent $i$ is the leader. If the condition couldn't be satisfied by any $e_i,i=1,2,\cdots,n$, $(\lambda I+J,P^{-1}e)$ will never be of full row rank, and the system is not controllable. $\square$

\begin{corollary}\label{slcud}
The following two assertions hold:\\
1. For multi-agent system (\ref{compact}), suppose that eigenvalues of the Laplacian matrix satisfy condition 1 of Theorem \ref{slc}. Agent $i$ can be selected as the single leader to control the system if and only if the $i$-th column in $P^{-1}$ satisfies condition 2 in Theorem \ref{slc}.\\
2. If multi-agent system (\ref{compact}) is SLC, there must be a spanning tree in the communication graph with the root being the leader.
\end{corollary}
Proof: Assertion 1 is a direct conclusion of Theorem \ref{slc} and the proof is omitted. For assertion 2, if the graph doesn't contain a spanning tree, then $rank(L)<n-1$, which will lead to that eigenvalue $0$ correspond to more than one Jordan blocks, and thus $L$ is not cyclic, i.e. system (\ref{compact}) is not controllable. $\square$

\begin{remark}
Although it is intuitive to judge controllability of multi-agent system (\ref{compact}) from the perspective of graph theory, to find a graphic necessary and sufficient condition for controllability is rather difficult. Ji et al. achieved a necessary and sufficient condition via an algebraic property of eigenvalues of the Laplacian matrix \cite{Ji09}, but the conclusion is only applicable to judge controllability of some specific systems with given leaders. However, Theorem \ref{slc} and Corollary \ref{slcud} showed necessary and sufficient conditions based on the Jordan form of $L$, which could not only judge controllability, but also solved SLC problem and contribute to searching for the fewest leaders. Jordan form of Laplacian matrix becomes a new channel to investigate controllability of multi-agent systems.
\end{remark}

\subsection{r leaders controllability}

Based on the SLC problem, a question arises that if system (\ref{compact}) is not SLC, wether how many leaders are needed at least to control the system? For a generic directed topology, the next theorem shows how to check controllability of system (\ref{compact}) with multiple leaders, as well as wether $|\mathbb{V}_l|$ is minimum.

For multi-agent system (\ref{compact}), $J=diag(J_0,J_1,\cdots,J_s)$ is the Jordan form of Laplacian matrix $L$. Distinct eigenvalues of $L$ are denoted as $\lambda_0,\lambda_1,\cdots,\lambda_t,t\leq s$. $P^{-1}LP=J$, $P^{-1}=(\eta^1,\eta^2,\cdots,\eta^n)$.
\begin{theorem}\label{mls}
System (\ref{compact}) is $r$ leaders controllable if and only if there exist $r$ columns in $P^{-1}$, denoted as $\bar{P}=(\eta^{c_1},\eta^{c_2},\cdots,\eta^{c_r})$, satisfying the following two conditions simultaneously:\\
1. Assume the geometric multiplicity of eigenvalue $\lambda_i$ is $k_i$, with the corresponding Jordan blocks $J_{i_1},J_{i_2},\cdots,J_{i_{k_i}}$, then $rank(\Omega_{\lambda_i})=k_i$, where
\begin{equation*}
  \Omega_{\lambda_i}=\left(
         \begin{array}{cccc}
           \eta^{c_1}_{m_{i_1}} & \eta^{c_2}_{m_{i_1}} & \cdots & \eta^{c_r}_{m_{i_1}} \\
           \eta^{c_1}_{m_{i_2}} & \eta^{c_2}_{m_{i_2}} & \cdots & \eta^{c_r}_{m_{i_2}} \\
           \vdots & \vdots & \ddots & \vdots \\
           \eta^{c_1}_{m_{i_{k_i}}} & \eta^{c_2}_{m_{i_{k_i}}} & \cdots & \eta^{c_r}_{m_{i_{k_i}}} \\
         \end{array}
       \right),
\end{equation*}
$i=0,1,2,\cdots,t$; $k_0+k_1+\cdots+k_t=s$;\\
2. Any combination of less than $r$ columns in $P^{-1}$ couldn't satisfy condition 1.\\
In this circumstance, agents $c_1,c_2,\cdots,c_r$ are able to control the system together, where $c_1,c_2,\cdots,c_r$ are also the column indices of $\bar{P}$ in $P^{-1}$.
\end{theorem}
Proof: Consider the matrix $(\lambda I+L,P^{-1}B)$, denote $\tilde{B}=P^{-1}B=(\tilde{B}_0^T,\tilde{B}_1^T,\cdots,\tilde{B}_s^T)^T$, where $\tilde{B}_i^T$ is of the same row size as $J_i$, $i=0,1,2,\cdots,s$.
\begin{equation}\label{pbh}
  \begin{array}{*{22}c}
    (\lambda I + J,\tilde{B})
  = (\left( {\begin{array}{*{20}{c}}
   {\lambda I + {J_0}} & {} & {} & {}  \\
   {} & {\lambda I + {J_1}} & {} & {}  \\
   {} & {} &  \ddots  & {}  \\
   {} & {} & {} & {\lambda I + {J_s}}  \\
\end{array}} \right),\left( {\begin{array}{*{20}{c}}
   {{\tilde{B}_0}}  \\
   {{\tilde{B}_1}}  \\
    \vdots   \\
   {{\tilde{B}_s}}  \\
\end{array}} \right)). \\
 \end{array}
\end{equation}
According to the PBH Test, system (\ref{compact}) is controllable if and only if rank of (\ref{pbh}) is $n$, i.e. all $(\lambda I+J_l,\tilde{B}_l),l=0,1,2,\cdots,s$ are of full row rank for any $\lambda\in\Lambda(-L)$. Consider the submatrix
\begin{equation*}
  (\left( {\begin{array}{*{20}{c}}
   {\lambda I + {J_{{i_1}}}} & {} & {} & {}  \\
   {} & {\lambda I + {J_{{i_2}}}} & {} & {}  \\
   {} & {} &  \ddots  & {}  \\
   {} & {} & {} & {\lambda I + {J_{{i_{{k_i}}}}}}  \\
\end{array}} \right),\left( {\begin{array}{*{20}{c}}
   {{\tilde{B}_{{i_1}}}}  \\
   {{\tilde{B}_{{i_2}}}}  \\
    \vdots   \\
   {{\tilde{B}_{{i_{{k_i}}}}}}  \\
\end{array}} \right)),
\end{equation*}
it is always of full row rank if and only if $rank(\Omega_{\lambda_i})=k_i$, therefore condition 1 is a necessary and sufficient for controllability of system (\ref{compact}). Condition 2 ensures the minimality of $\mathbb{V}_l$. $\square$

If only condition 1 of Theorem \ref{mls} is satisfied, the system is also controllable, whereas $|\mathbb{V}_l|$ may not be minimal.
\begin{corollary}
Assume the geometric multiplicity of eigenvalues $\lambda_0,\lambda_1,\lambda_2,\cdots,\lambda_t$ are $k_0,k_1,k_2,\cdots,k_t$ respectively, denote $k=\max\{k_0,k_1,k_2,\cdots,k_t\}$, then the minimum $|\mathbb{V}_l|=r$ satisfies $k\leq r \leq \sum\limits_{l=0}^t k_l$.
\end{corollary}
Proof: Without loss of generality, suppose $k=k_1$. On the one hand, if $|\mathbb{V}_l|<k$, $\Omega_{\lambda_1}$ is not of full row rank, i.e. $rank(\Omega_{\lambda_1})<k_1$, which contradicts condition 1 in Theorem \ref{mls}. On the other hand, there always exist $k_i$ linearly independent columns in $P^{-1}$ to ensure $rank(\Omega_{\lambda_i})=k_i$, therefore $|\mathbb{V}_l| \leq \sum\limits_{l=0}^s k_l$. $\square$

Be worth mentioning, $k$ leaders are not always enough to control the whole system, see Example (\ref{exa4}). Theorem \ref{mls} is just a theoretical result, not a direct method to search for the fewest leaders. For a generic directed graph, this problem is extremely similar to the minimal controllability problem proposed in \cite{Olshevsky14}, which appears to be NP-hard. How to put forward an effective algorithm to search for the fewest leaders is a problem worthy of study.

\section{Weight adjustment problem}

As introduced, there are two methods to improve controllability of a multi-agent system, one is by adding leaders, and the other is by adjusting edge weights. When the leaders are fixed, controllability could be achieved only by assigning new weights to some proper edges. This yields weight adjustment problem, which will be discussed in this section.
\subsection{Structural controllability}

In order to control a system by adjusting edge weights, the system should be structurally controllable. We start investigating structural controllability from a property of tree graphs.
\begin{proposition}\label{tree}
$\mathbb{T}_{v}$ with root $v$ as the single leader is controllable if and only if edges in different branches in $\mathbb{T}_v$ have no equal weights.
\end{proposition}
Proof: This conclusion is first achieved in \cite{Guan}. Here we show an extra proof in a pure algebraic method, which will benefit comprehending proof of Theorem \ref{wa} in the following. See the Appendix. $\square$

\begin{theorem}\label{slwa}
System (\ref{compact}) is structurally controllable with one leader if and only if the communication graph contains a spanning tree with the root being the leader.
\end{theorem}
Proof: (Necessity) If the communication graph doesn't contain a spanning tree, there must be at least two agents that couldn't get information from each other, and thus the minimum spanning forest contains more than one trees, denoted as $\mathbb{T}_1,\mathbb{T}_2,\cdots,\mathbb{T}_r$. Once the leader is selected in some $\mathbb{T}_i$, there always be agents in other trees that couldn't get information from the leader and apparently the system is not controllable.\\
(Sufficiency) Suppose the Laplacian matrix of the spanning tree $\mathbb{T}$ is $L_T$, and the corresponding similarity transformation matrix is $P_T$. Consider the communication topology $\mathbb{G}$ with the Laplacian matrix $L=L_T+\epsilon L_R$, where $L_R$ is the Laplacian matrix of the subgraph of $\mathbb{G}$ by deleting the edges in $\mathbb{T}$. Suppose the similarity transformation matrix of $L$ is $P$, $\Delta P\triangleq P^{-1}-P^{-1}_T$. Apparently, when $\epsilon\rightarrow 0$, $\Delta P\rightarrow 0$. There exists a combination of weights that makes all eigenvalues of $L_T$ distinct and $\mathbb{T}$ is controllable by Proposition \ref{tree}. Therefore, all entries in the first column of $P^{-1}_T$ are not $0$. When $\epsilon$ is small enough, all eigenvalues of $L$ remain distinct and all entries in $\Delta P$ will be small enough such that the first column of $P^{-1}$ contains no $0$. Therefore, system (\ref{compact}) is structurally controllable with the root of $\mathbb{T}$ being the single leader. $\square$

\begin{corollary}\label{mlwa}
System (\ref{compact}) is structurally controllable with at least $r$ leaders if and only if the minimum spanning forest of communication graph contains $r$ trees with the roots being the leaders.
\end{corollary}
Proof: (Necessity) Refer to the necessity proof of Theorem \ref{slwa}.\\
(Sufficiency) When $r=2$, i.e. the spanning forest $\mathbb{F}=\{\mathbb{T}_{v_1},\mathbb{T}_{v_2}\}$. Since the graph is not structurally controllable, at least two leaders are needed. Select $v_1$ and $v_2$ as leaders. With proper weights, $\mathbb{T}_{v_1}$ and $\mathbb{T}_{v_2}$ could be controlled by $v_1$ and $v_2$ respectively. For each edge whose parent node lies in $\mathbb{T}_{v_1}$ (or $\mathbb{T}_{v_2}$) and the child node lies in $\mathbb{T}_{v_2}$ (or $\mathbb{T}_{v_1}$), assign a weight small enough to neglect the effect of it, then the whole graph remains controllable. Hence the conclusion holds for $r=2$. Suppose the conclusion holds for $r=n$. When $r=n+1$, i.e. $\mathbb{F}=\{\mathbb{T}_{v_1},\mathbb{T}_{v_2},\cdots,\mathbb{T}_{v_{n+1}}\}$, by the induction hypothesis, any $n$ trees in $\mathbb{F}$ are structurally controllable with their roots being the leaders. Without loss of generality, suppose $\mathbb{T}_{v_1},\mathbb{T}_{v_2},\cdots,\mathbb{T}_{v_n}$ are controllable. Select $v_{n+1}$ as a new leader, assign proper weights to $\mathbb{T}_{n+1}$, $\mathbb{F}$ is controllable. Assign small weights to the connections among the trees could make the whole graph controllable. According to mathematical induction, the conclusion holds for any positive integer $r$. $\square$

As introduced, in the existing results, structural controllability of multi-agent systems is studied based on absolute protocols, due to the $0$ row sum limitation. In Theorem \ref{slwa} and Corollary \ref{mlwa}, we showed the graphic conditions for the distributed protocol. These two conclusions are consistent to the leader-follower connected structure \cite{Ji09}, which is also a necessary and sufficient condition for absolute protocols.

\subsection{Fewest edges to be assigned new weights}

For a structurally controllable system, a problem arise afterwards that how to adjust weights on fewest edges to achieve controllability. The next theorem will put forward the number of the fewest edges. To describe more explicitly, the problem of weight adjustment problem is defined mathematically as follow.
\begin{problem}\label{wrp}
$\mathbf{Weight~adjustment~problem}$: For multi-agent system (\ref{compact}), when it is not controllable whereas all the roots in the spanning forest of the interaction topology are chosen as leader, find a set of edges $\mathbb{E}_m\subseteq\mathbb{E}$ with minimum $|\mathbb{E}_m|$, such that when the weights are properly adjusted on edges in $\mathbb{E}_m$, the system is controllable without changing the leader.
\end{problem}
Investigating weight adjustment problem first requires that, when taking all the leaders as roots, there must be a forest with these roots covering all nodes in the interaction graph. Especially, when there is only one leader, the graph should contain a spanning tree.
\begin{theorem}\label{wa}
Suppose that the communication graph of multi-agent system (\ref{compact}) contains a directed spanning tree, and the root is the single leader. If the rank of controllability matrix is $n-r$, then there exist $r$ edges such that the system could be controllable by adjusting weights on them, and any adjustment on less than $r$ edges cannot make system (\ref{compact}) controllable.
\end{theorem}
Proof: When $r=0$, the conclusion is obvious. Without loss of generality, assume the agents are labeled as follow: Label the root as $1$, get the distance partition $\{D_0,D_1,\cdots,D_p\}$ where $D_0$ is the root, and there is no $D_\infty$ due to the existence of spanning tree in the graph. Label the nodes from those in $D_1$ to those in $D_p$ successively. With this method, for agent $i$ in $D_q,q=2,3,\cdots,p$, the first $q-1$ entries of the $i$-th row in $L$ are $0$.\\
For $1\leq r<n$, we prove that there exists one edge whose weight if be adjusted properly, could increase the rank of controllability matrix $C$ by $1$. Mathematically, this equals to prove that if $k_i\neq0,i=1,2,\cdots,s,$
\begin{equation}\label{linear1}
  (k_1e_{i_1}+k_2e_{i_2}+\cdots+k_se_{i_s})^TL^me_1=0,
\end{equation}
for $m=0,1,2,\cdots,n-1$, then, there exist $\Delta L$ and $m_0$ such that $(k_1e_{i_1}+k_2e_{i_2}+\cdots+k_se_{i_s})^T(L+\Delta L)^{m_0}e_1\neq0$ where $\Delta L$ contains only two opposite nonzero elements who lie in a same row. The positive one is in the principal diagonal and the other is in front of it, $m_0\leq n-1$. \\
Suppose the two nonzero elements are in the $j$-th row, $j\neq i_t, t=1,2,\cdots,s$, it can be verified,
\begin{equation}\label{linear2}
  (k_1e_{i_1}+k_2e_{i_2}+\cdots+k_se_{i_s})^T(L+\Delta L)^me_1=0
\end{equation}
for all $m=0,1,2,\cdots,n-1$. If we intend to increase rank of $C$, the revised edge must be selected from one of the $i_1,i_2,\cdots,i_s$-th rows in $L$. Take the $i_1$-th row as an example. \\
Next we show the existence of $\Delta L$ and $m_0$. Suppose that there is an edge from agent $i_1$ to agent $j$, i.e. $L_{j,i_1}\neq 0$, and the equation (\ref{linear2}) holds for $m=0,1,2,\cdots,n-1$, $\epsilon>0$. Here $\Delta L_{j,i_1}=-\epsilon, \Delta L_{i_1,i_1}=\epsilon$ and the other entries in $\Delta L$ are all $0$. Under the assumption (\ref{linear1}), combined with (\ref{linear2}), we get
\begin{equation}\label{linear3}
  (k_1e_{i_1}+k_2e_{i_2}+\cdots+k_se_{i_s})^T((L+\Delta L)^m-L^m)e_1=0
\end{equation}
for $m=0,1,2,\cdots,n-1$. When $m=1$, $(k_1e_{i_1}+k_2e_{i_2}+\cdots+k_se_{i_s})^T\Delta L e_1=-k_1\Delta L_{i_1,1}=0$ yields $\Delta L_{i_1,1}=0$, which means the $L_{i_1,1}=0$ and thus agent $i_1$ couldn't get information from the root of the spanning tree. Denote $D_m=(L+\Delta L)^m-L^m$, therefore $D_{m+1}=D_mL+D_m\Delta L+L^m\Delta L$. Since the $(i_1,1)$ entry in $D_m\Delta L$ and $L^m\Delta L$ are both $0$, the $(i_1,1)$ entry in $D_m L$ must also be $0$ to satisfy (\ref{linear3}). Considering that the $(i_1,1)$ entry in $D_m L$ is $\Delta L_{i_1i_1}-\Delta L_{i_1,j}$, and this will lead to $L_{i_1i_1}=L_{i_1,j}$, which contradicts the fact that $L_{i_1i_1}>0$ and $L_{i_1,j}<0$. Here we get that there exist $\Delta L$ and $m_0$ such that $(k_1e_{i_1}+k_2e_{i_2}+\cdots+k_se_{i_s})^T(L+\Delta L)^{m_0}e_1\neq0$.\\
Apparently, $\Delta L$ only changes the $i_1$-th row of $C$. Without loss of generality, suppose no $s-1$ vectors of $C_{r_{i_1}},C_{r_{i_2}},\cdots,C_{r_{i_s}}$ are linearly dependent. Next we prove that there exists a proper $\epsilon>0$ such that $C_{r_{i_1}}+\delta C_{r_{i_1}},C_{r_{i_2}},\cdots,C_{r_{i_s}}$ are linearly independent. Consider the equation $k_1^{'}(C_{r_{i_1}}+\delta C_{r_{i_1}})+k_2^{'}C_{r_{i_2}}+\cdots+k_s^{'}C_{r_{i_s}}=0$, it is equal to $(k_1^{'}-k_1)C_{r_{i_1}}+(k_2^{'}-k_2)C_{r_{i_2}}+\cdots+(k_s^{'}-k_s)C_{r_{i_s}}+k_1^{'}\delta C_{r_{i_1}}=0$. With the discussion above, we can get that if $\delta C_{r_{i_1}}\neq0$, $\delta C_{r_{i_1}}$ will change with $\epsilon$ nonlinearly and thus a proper $\epsilon$ will ensure that $\delta C_{r_{i_1}}$ is linearly independent to $C_{r_{i_1}}+\delta C_{r_{i_1}},C_{r_{i_2}},\cdots,C_{r_{i_s}}$. If $k_1^{'}\neq0$, $(k_1^{'}-k_1)C_{r_{i_1}}+(k_2^{'}-k_2)C_{r_{i_2}}+\cdots+(k_s^{'}-k_s)C_{r_{i_s}}+k_1^{'}\delta C_{r_{i_1}}$ will never be $0$, therefore $k_1^{'}=0$. Since $C_{r_{i_1}}+\delta C_{r_{i_1}},C_{r_{i_2}},\cdots,C_{r_{i_s}}$ are linearly independent, once $k_1^{'}=0$, $k_i^{'}-k_i=-k_i$ for all $i=2,3,\cdots,s$. Finally, $k_1^{'}=k_2^{'}=\cdots=k_s^{'}=0$. This means that, there exist proper $\Delta L$ and $m_0$ to eliminate one of the linearly dependent rows in the controllability matrix. Based on this, assigning a proper weight to one proper edge, the rank of $C$ will increase $1$ and only $1$. This implies that exactly $r$ different edges from different rows are needed to be adjusted to fulfill the decreased rank of $C$. $\square$

\begin{remark}\label{war}
Refer to the proof of Theorem \ref{wa}, the next two conclusions can be achieved.\\
1. Suppose that controllability of system (\ref{compact}) can be improved by adjusting the weight on edge $e^*$ and the weight increment is $\epsilon$, i.e. $\Delta L_{e^*}=-\epsilon$, then there exists an $n-1$ order polynomial of $\epsilon$, say $f(\epsilon)$, such that $\Delta L_{e^*}$ fails to increase rank of $L$ if and only if $f(\epsilon)=0$. Hence, there are no more than $n-1$ values of $\epsilon$ that would fail to improve controllability. Therefore, if we randomly endue a new weight to $e^*$, the probability of successfully increase the rank of $C$ is $1$.\\
2. If $e^*$ should be selected from the $i$-th row of $L$, denoted as $L_{r_i}$, then the first nonzero entry in $L_{r_i}$ could be the edge to adjust weight. This means we could redesign the weight of the edge that connects to agent $i$ from the agent with the minimum identifer in $N_i$.
\end{remark}

Next we show an algorithm on how to perform weight adjustment on proper edges. To express more explicitly, the algorithm here is designed for the graph that contains a spanning tree. However, it can be improved to fit for generic directed graphs.
\begin{table}
\begin{tabular}{lccc}
\toprule
   \textbf{Algorithm} \\
   \midrule
Get all the nodes that could be a root of a spanning tree, identify them as $v_1,v_2,\cdots,v_m$, $C=\mathbf{0}_{n\times n}$;\\
Select an weight increment $\theta>0$.\\
$\mathbf{For}$ $k=1:m$\\
~~~~Label the root $v_k$ as $1$, get the distance partition $\{D_0={v_k},D_1,\cdots,D_p\}$, label the whole system\\
~~~~from nodes in $D_1$ to nodes in $D_p$ successively, get the Laplacian matrix $L$ and the controllability matrix $C$;\\
~~~~$\mathbf{if}$ $rank(C)==n$\\
~~~~~~~~Output ``The system is controllable with leader agent $k$'', exit;\\
~~~~$\mathbf{else~if}$ $rank(C)<r$\\
~~~~~~~~$r=rank(C),s=k$;\\
~~~~$\mathbf{end~if}$\\
$\mathbf{end~for}$\\
Use row elimination to get the all-$0$ rows, get their row identifers $i_1,i_2,\cdots,i_s$;\\
$\mathbf{while}$ $rank(C)<n$\\
~~~~$\mathbf{for}$ $1\leq j\leq s$\\
~~~~~~~~Add weight $j\theta$ to the edge corresponding to the first nonzero element in the $i_j$-th row of $L$;\\
~~~~$\mathbf{end~for}$\\
~~~~Get $\tilde{L}, L=\tilde{L}$; Calculate $C$; $\theta=1.1\theta$;\\
$\mathbf{end~while}$\\
Output ``The number of fewest edges to be assigned new weights is $s$, and an available graph Laplacian is $L$''.\\
   \bottomrule
\end{tabular}
\end{table}

\section{An application}

In this subsection, an application of controllability and structural controllability will be shown under a kind of special graphs named in-degree regular graphs.
\begin{definition}\label{indegree}
In-degree Regular Graph: A directed graph is called in-degree regular if the in-degrees of each node are equal, i.e. $\deg_{in}(i)=\deg_{in}(j)$ for all $1\leq i,j\leq n$.
\end{definition}

\begin{theorem}\label{in-degree}
An in-degree regular graph can be controlled by agent $1$ if and only if matrix $M=[m_{ij}]\in \mathbb{R}^{(n-1)\times (n-1)}$ is invertible, where $m_{ij}$ is the number of different paths from agent $1$ to agent $i+1$ with length $j$.
\end{theorem}
Proof: According to Definition \ref{indegree}, the Laplacian matrix $L=D-A=dI-A$, $L^k=\sum\limits^k_{i=0} C_k^i d^i(-A)^{n-i}$. The controllability matrix $C=(e,Le,L^2e,\cdots,L^{n-1}e)$, and
\begin{eqnarray*}
   {rank(C)}={rank(e,(dI - A)e, \cdots ,\sum\limits_{i = 0}^{n - 1} {C_k^i } d^i ( - A)^{n - i} e)} ={rank(e,Ae, \cdots ,A^{n - 1} e)}.
\end{eqnarray*}
As assumed, $e=(1,0,\cdots,0)^T$ and the $i$-th entry of $A^je$ is the number of paths from agent $1$ to agent $i$ with length $j$. $M$ is the submatrix of $C$ by deleting the first row and the first column. When $M$ is invertible, $C$ is of full rank and thus the system is controllable, vice versa. $\square$

For in-degree regular graphs, controllability can be validated more intuitively.
\begin{corollary}
The next two assertions on in-degree regular graphs hold.\\
1. For an in-degree regular graph with $n$-nodes, whose adjacency matrix is $A$. If each column in $\sum\limits^{n-1}_{i=1} A^i$ contains at least one $0$, the system is not SLC.\\
2. For an in-degree regular graph, denote $S=\sum\limits_{k=1}^{n-1} A^k$, if at least $m$ columns of $S$ are needed to ensure the sum of them contains no $0$ entry, then it should be $|\mathbb{V}_l|\geq m$ to make system (\ref{compact}) controllable.
\end{corollary}
Proof: For assertion 1, if each column of $\sum\limits^{n-1}_{i=1} A^i$ contains at least one $0$, no matter which agent is selected as the leader, there will be at least one agent that couldn't get information from leader, and this makes an in-degree regular graph uncontrollable. \\
For assertion 2, choosing $m$ columns of $S$ whose sum contains no $0$ entry is to ensure a leader-follower connected structure. Therefore, at least $m$ leaders are needed for controllability. $\square$
\begin{proposition}
An in-degree regular graph is structurally controllable if and only if there exists one column of $\sum\limits_{k=1}^{n-1} A^k$ that contains no $0$ entry except for the principal diagonal elements, where $A$ is the adjacency matrix of the graph.
\end{proposition}
Proof: Without loss of generality, consider the first column of $\sum\limits_{k=1}^{n-1} A^k$, delete the first entry and denote the remained vector as $\eta$. $\eta_i\neq 0$ if and only if there exists a path from agent $1$ to agent $i+1$. Therefore, $\eta$ contains no $0$ entry if and only if the graph contains a spanning tree, which is a necessary and sufficient condition of structural controllability. $\square$

\section{Simulation}

Two numerical examples are presented in this section to illustrate the effectiveness of theoretical results.

\begin{figure}[!t]
\centering
\includegraphics[width=2.5in]{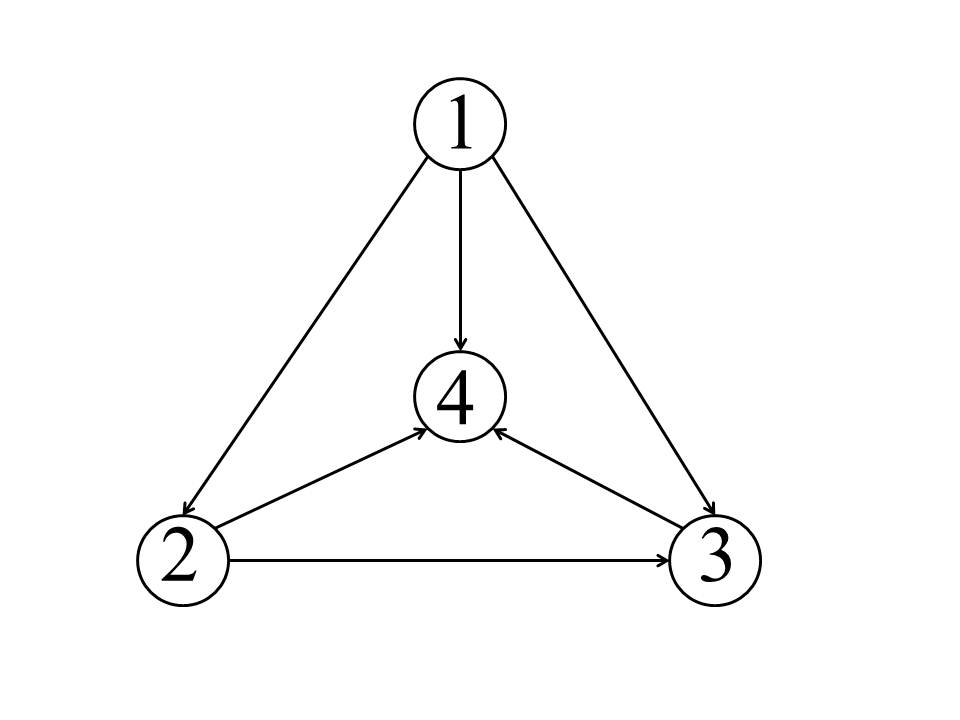}
\caption{Interaction topology of Example \ref{exa4}}
\label{fig4}
\end{figure}

\begin{example}\label{exa4}
Figure \ref{fig4} shows a system with four agents. The Laplacian matrix and the corresponding $P^{-1}$ are shown as follow:
{\small
\begin{equation*}
  L=\left(
          \begin{array}{cccc}
          0 & 0 & 0 & 0 \\
          -1 & 1 & 0 & 0 \\
          -1 & -1 & 2 & 0 \\
          -1 & -1 & -1 & 3 \\
          \end{array}
          \right),
  P^{-1}=\left(
                                                          \begin{array}{cccc}
                                                            1 & 0 & 0 & 0 \\
                                                            -1 & 1 & 0 & 0 \\
                                                            0 & -1 & 1 & 0 \\
                                                            0 & 0 & -1 & 1 \\
                                                          \end{array}
                                                        \right).
\end{equation*}
}
There are at least two $0$ in each column of $P^{-1}$, therefore Figure \ref{fig4} is not SLC. If agent $1$ and agent $3$ are selected as leaders, the graph is controllable, but once agent $2$ is selected as a leader, two more leaders are needed. Moreover, it also demonstrates that even if the eigenvalues of the Laplacian matrix are distinct, Figure \ref{fig4} may also not be SLC.
\end{example}

\begin{figure}[!t]
\centering
\includegraphics[width=2.5in]{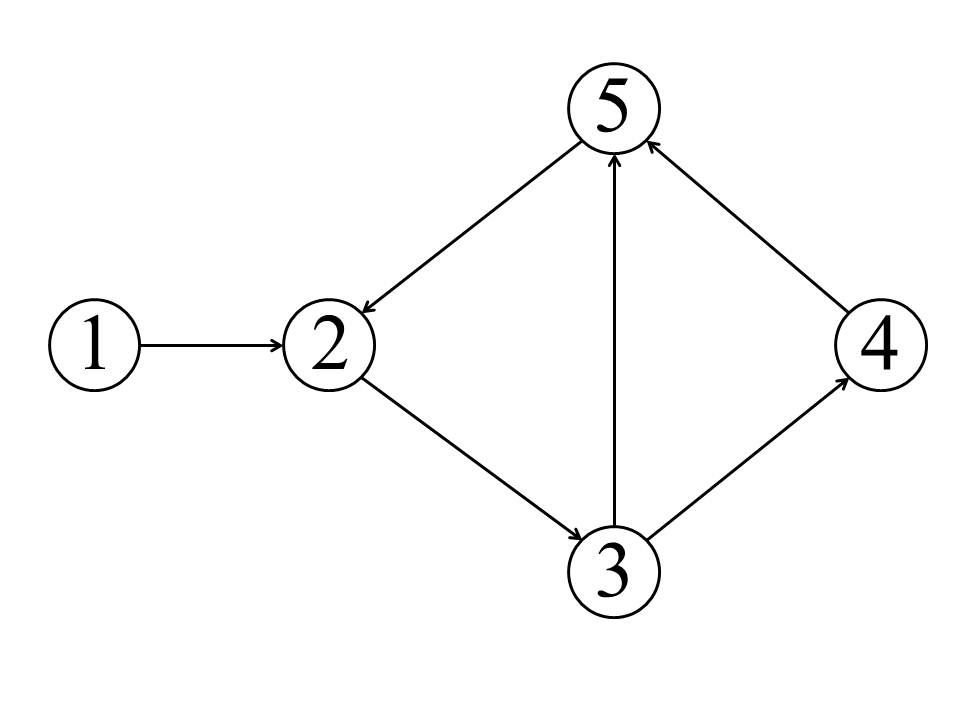}
\caption{Interaction topology of Example \ref{exa6}}
\label{fig6}
\end{figure}

\begin{example}\label{exa6}
Figure \ref{fig6} shows a directed communication graph of system (\ref{compact}). The Laplacian matrix is\\
{\small
\begin{equation*}
  L=\left(
                                                                    \begin{array}{ccccc}
                                                                      0 & 0 & 0 & 0 & 0 \\
                                                                      -1 & 2 & 0 & 0 & -1 \\
                                                                      0 & -1 & 1 & 0 & 0 \\
                                                                      0 & 0 & -1 & 1 & 0 \\
                                                                      0 & 0 & -1 & -1 & 2 \\
                                                                    \end{array}
                                                                  \right)
\end{equation*}
}
with eigenvalues $0,2,0.2451,1.8774 \pm 0.7449i$, and
{\small
\begin{equation*}
  P^{-1}=\left(
                                                                   \begin{array}{ccccc}
                                                                     1 & 0 & 0 & 0 & 0 \\
                                                                     0 & 0 & 0 & 1 & -1 \\
                                                                     -1.2672 & 0.3106 & 0.5451 & 0.2345 & 0.1770 \\
                                                                     0.1336 + 0.1283i & -0.1553 - 0.3404i & -0.2726 + 0.0740i & -0.1172 + 0.4143i & 0.4115 - 0.2762i \\
                                                                     0.1336 - 0.1283i & -0.1553 + 0.3404i & -0.2726 - 0.0740i & -0.1172 - 0.4143i & 0.4115 + 0.2762i \\
                                                                   \end{array}
                                                                 \right).
\end{equation*}
}
Since there is at least one $0$ in each column, the system is not SLC. Actually it can be controlled by two leaders, one of which must be agent $1$ and the other could be agent $4$ or $5$. Meanwhile, the graph contains a spanning tree, hence system (\ref{compact}) is structurally controllable with one leader. Select agent $1$ as the leader, we can get that rank of the controllability matrix is $4$, thus the system can be controlled only adjusting the weight on one edge. Here revise $w_{35}$ from the original $1$ to $1.1$, and the Laplacian matrix turns to be
{\small
\begin{equation*}
  \tilde{L}=\left(
                                                                    \begin{array}{ccccc}
                                                                      0 & 0 & 0 & 0 & 0 \\
                                                                      -1 & 2 & 0 & 0 & -1 \\
                                                                      0 & -1 & 1 & 0 & 0 \\
                                                                      0 & 0 & -1 & 1 & 0 \\
                                                                      0 & 0 & -1.1 & -1 & 2.1 \\
                                                                    \end{array}
                                                                  \right),
\end{equation*}
}
with the eigenvalues $0,0.2493,1.8930,1.9788 \pm 0.7305i$, and the first column of the corresponding $\tilde{P}^{-1}$ is $[1,-1.2639, 0.0105,0.1267 + 0.1412i,0.1267 - 0.1412i]^T$, which mean the system becomes controllable. This illustrates the conclusions in Section 4.
\end{example}

\section{Conclusion}

This paper has studied controllability of multi-agent systems with directed communication topologies. The concept of leader selection was explicitly introduced. Algebraic necessary and sufficient conditions on how to select the fewest leaders were presented based on the Jordan blocks of $L$ and the corresponding transformation matrix $P^{-1}$. Considering that controllability may also be achieved by adjusting edge weights, this paper also studied the weight adjustment problem, which aims to determine the fewest edges to be assigned new weights to ensure controllability, as well as the new weights. The result showed that a multi-agent system is structurally controllable if and only if the communication graph contains a spanning tree. The number of fewest edges equals the rank deficiency of controllability matrix. An algorithms on how to perform wight adjustment was also provided.

\section{Appendix}
Proof of Proposition \ref{tree}\\
(Necessity) If two edges of $\mathbb{T}_v$ in different branches share a common weight, there must be two equal elements $L_{ii}=L_{jj}=\lambda$ in the principal diagonal of $L$, and $L_{ij}=L_{ji}=0$. Since $\lambda$ is an eigenvalue of $L$, there must be two linearly independent eigenvectors of $\lambda$. Therefore, in the Jordan form of $L$, two Jordan blocks share a common eigenvalue $\lambda$. According to Theorem \ref{slc}, the system is not SLC.\\
(Sufficiency) For clarity, the proof is stated in four steps. First assume all the edges have different weights, hence $P^{-1}LP=D$ where $D$ is the diagonal form of $L$, and we prove that $p_{ij}=0$ for all $i<j$. Then we show the expressions of $p_{ij}$ for all $1\leq i,j\leq n$. Next we prove that none of the elements in the first column of $P^{-1}$ is $0$. Finally we prove that controllability will not be broken when adding one leaf to a controllable tree with any weight different from weights in other branches.\\
Part 1: Denote $R=\lambda_jI-L$, obviously when $j\geq1$, the only nonzero element in the first row of $R$ is $r_{11}$, thus $p_{1j}=0$. Suppose when $i< i^*$, where $i^*\leq j-1$, $p_{ij}=0$, since $0=\sum\limits_{k=1}^n l_{i^*k}p_{kj}-\lambda_j p_{jj}=\sum\limits_{k=1}^{i^*} l_{i^*k}p_{kj}=l_{i^*i^*}p_{i^*j}$ and $l_{i^*i^*}\neq 0$, so $p_{i^*j}=0$. According to mathematical induction, $p_{ij}=0$ for all $i<j$.\\
Part 2: Since $p_i=(p_{1i},p_{2i},\cdots,p_{ni})^T$ is an eigenvector of $L$, $(L-\lambda_iI)p_i=0$, which means $\sum\limits_{k=1}^n l_{ik}p_{ki}-\lambda_i p_{ii}=0$. From part 1 we know $p_{1i}=p_{2i}=\cdots=p_{i-1,i}=0$, and for $\mathbb{T}_v$, $l_{i,i+1}=l_{i,i+2}=\cdots=l_{i,n}=0$, this yields $(l_{ii}-\lambda_i)p_{ii}=0$. Owing to $l_{ii}-\lambda_i=0$, $p_{ii}$ could be any number, and $p_{ii}=1$ is chosen here without loss of generality. For each $i>j$, we get $0=\sum\limits_{k=1}^n l_{ik}p_{ki}-\lambda_j p_{ij}=\sum\limits_{k=1}^i l_{ik}p_{ki}-\lambda_j p_{ij}$. As mentioned before, there exists one and only one $k_i<i$ for each $i$ such that $l_{ik}\neq 0$. Correspondingly, $p_{ij}=\frac{l_{ik_i}p_{k_ij}}{\lambda_j-l_{ii}}$. Combine the results afore yields\\
\begin{equation*}
  p_{ij}=\left\{
          \begin{array}{cc}
            0, & i<j \\
            1, & i=j \\
            \frac{l_{ik_i}p_{k_ij}}{\lambda_j-l_{ii}}, & i>j
          \end{array}
        \right.
\end{equation*}
Part 3: Now consider the first column of $P^{-1}$, denoted as $q=(q_1,q_2,\cdots,q_n)^T$. Obviously, $q_1\neq0$, otherwise, $Pq\neq e$ where $e=(1,0,0,\cdots,0)^T$. Suppose $q_i \neq 0$ with $i<i^*$, and $\sum\limits_{k=1}^{i} p_{ik}q_k=e_i$. Since $e_{i^*}=\sum\limits_{k=1}^n p_{i^*k}q_k=\sum\limits_{k=1}^{i^*} p_{i^*k}q_k$, $q_{i^*}=-\sum\limits_{k=1}^{i^*-1}p_{i^*k}q_k=-l_{i^*k_{i^*}}
(\frac{p_{k_{i^*}1}}{\lambda_1-l_{i^*i^*}},\frac{p_{k_{i^*2}}}{\lambda_2-l_{i^*i^*}},\cdots,
\frac{p_{k_{i^*},i^*-1}}{\lambda_{i^*-1}-l_{i^*i^*}})(q_1,q_2,\cdots,q_{i^*-1})^T$. It follows from part 2 that the only nonzero element in the $i^*$ row of $P$ except for $p_{i^*1}=p_{i^*i^*}=1$ is $p_{i^*k_{i^*}}$. Denote $\xi_{i^*}=-l_{i^*k_{i^*}}(\frac{p_{k_{i^*}1}}{\lambda_1-l_{i^*i^*}},\frac{p_{k_{i^*2}}}{\lambda_2-l_{i^*i^*}},
\cdots,\frac{p_{k_{i^*},i^*-1}}{\lambda_{i^*-1}-l_{i^*i^*}})^T$, and $\eta_j=(p_{j1},p_{j2},\cdots,p_{j,i^*-1})^T$. Obviously, $\eta_j$ are linear independent, $i=1,2,\cdots,i^*-1$. Thus, there exist $c_1,c_2,\cdots,c_{i^*-1}$ such that $\sum\limits_{k=1}^{i^*-1} c_k\eta_k^T=\xi_i^T$. Consider $\eta_2,\cdots,\eta_{i^*-1},\xi$, which are also linearly independent, so that $c_1\neq 0$. According to the induction hypothesis, $q_{i^*}=(q_1,q_2,\cdots,q_{i^*-1})\sum\limits_{k=1}^{i^*-1} c_k\eta_k=\sum\limits_{k=1}^{i^*-1} c_ke_k=c_1\neq 0$. Thus, each element in $q$ is not $0$, and in this case, condition 2 in Theorem \ref{slc} is satisfied.\\
Part 4: Suppose the diagonal form of $L$ is $J$, $P^{-1}LP=J$. If weight $\mu$ of the new edge is different from every other weight, the Jordan form of the new graph is
\begin{equation*}
  \hat{J}=\left(
            \begin{array}{cc}
              J &  \\
               & \mu \\
            \end{array}
          \right)=\left(
                    \begin{array}{cc}
                      P^{-1} & 0 \\
                      \alpha^T & 1 \\
                    \end{array}
                  \right)\left(
                           \begin{array}{cc}
                             L & 0 \\
                             \gamma^T & \mu \\
                           \end{array}
                         \right)\left(
                                  \begin{array}{cc}
                                    P & 0 \\
                                    \beta^T & 1 \\
                                  \end{array}
                                \right).
\end{equation*}
This means $\alpha^TP+\beta^T=0$ and $\alpha^TLP+\gamma^TP+\mu\beta^T=\beta^T(\mu I-J)+\gamma^TP=0$. It follows that
\begin{equation}\label{dirtree}
  \gamma^T=\alpha^T(\mu I-L).
\end{equation}
According to the proper of $\gamma$ and $L$ in (\ref{dirtree}), we can easily prove that all entries in $\alpha$ that correspond to the path from root to the new leaf are not $0$. It follows from Theorem \ref{slc} that the system is controllable. On the other hand, if $\mu$ equals to the weight of an edge in the path from the root to the new edge, $\mu$ is different from any weights in other branches due to the necessity conclusion. This guarantees that there is only one eigenvector $c(0,0,\cdots,0,1)^T$ correspond to eigenvalue $\mu$, where $c$ is a nonzero coefficient, and thus condition 1 in Theorem \ref{slc} is satisfied. Condition 2 can be proved similarly. $\square$

\end{document}